\documentclass[aps,prl,reprint,superscriptaddress,nofootinbib,floatfix]{revtex4-2}
\usepackage{graphicx}
\usepackage{amsmath,amssymb}
\usepackage{bm}

\begin{document}

\title{Dynamical phase selection controls compute scaling in looped transformers}

\author{Gunn Kim}
\thanks{Corresponding author: gunnkim@sejong.ac.kr}
\affiliation{Department of Physics, Sejong University, Seoul 05006, Republic of Korea}

\date{\today}

\begin{abstract}
A looped transformer performs inference by iterating a weight-tied map, making its computation a dynamical process whose cost is set by the resulting inference dynamics. Here we show that networks with identical architecture and objective, trained to identical accuracy, nevertheless realize distinct dynamical phases depending strongly on initialization, and that the bifurcation defining each phase determines how test-time compute scales. The phases are distinguished by their bifurcation mechanisms, including a saddle-node fold and a Neimark–Sacker-type transition to bounded nonstationary motion. In the fold phase, a one-dimensional normal-form reduction predicts both the relaxation-time and spectral-gap amplitudes from local derivatives of the trained map, yielding the parameter-free relation $\tau(\varepsilon)[1-\lambda_{\max}(-\varepsilon)]\to\pi$. Composed with a regular distribution of problem difficulty, the same critical slowing down produces the workload-level tail $P(\tau>N)\sim N^{-2}$. In the Neimark--Sacker phase, the fold scaling law disappears rather than merely changing its prefactor. Thus, test-time compute is not determined by architecture alone. It is governed by the dynamical phase of the solution found by training.
\end{abstract}

\maketitle

\textit{Introduction.}---A looped transformer applies one weight-tied block recurrently to arbitrary depth~\cite{dehghani2019,giannou2023,saunshi2025}, an architecture now scaled into language models that convert test-time iteration into reasoning performance~\cite{geiping2025,jeddi2026,movahedi2026}. Stripping the model to its recurrence makes inference explicit,
\begin{equation}
z_{t+1}=F_\theta(z_t;h),
\end{equation}
with the iteration count an operational relaxation time. This raises a basic question: \emph{what determines how long the computation takes?} Two aspects have received attention: what training makes the network compute, and whether the recurrence remains stable, often through spectral conditions on its linearization~\cite{prairie2026,buehlmaier2026,movahedi2026,labovich2026}. Neither fixes the cost of inference, which is set by the fixed-point structure the network relaxes toward and the bifurcation through which a decision is made.

That the linear condition is not the whole story is known: a spectral bound $\rho(\overline{\bm A})<1$ on the linear part does not certify convergence of the full map, and a minority of randomly initialized instances circulate on bounded orbits without settling~\cite{buehlmaier2026}. Our networks satisfy $\rho(\overline{\bm A})=0.90<1$ by construction and nonetheless have a leading multiplier of the full Jacobian that reaches or crosses unity at a finite field. We show that these behaviors are not failures but distinct dynamical phases selected by training, and that the phase determines the scaling law of inference time.

\textit{Setup.}---We iterate a weight-tied pre-norm attention-plus-MLP block on a latent state $z\in\mathbb{R}^{d}$ with one input token carrying the evidence $h$, residual scale $\rho$, and latent decay $\gamma$. The decay permits isolated fixed points: without it, normalization leaves fewer effective degrees of freedom than the number of fixed-point conditions, making an isolated solution nongeneric~\cite{SM}. Networks are trained end to end with a stationarity penalty to decide $\mathrm{sign}(h)$ from evidence withdrawn mid-episode in a fraction $\mu$ of trials, so the decision must be maintained without support. The objective specifies neither a fixed point nor a bifurcation nor an iteration count. We sweep $d=16$, $K=48$, $\mu\in\{0.20,\ldots,0.80\}$, and three initializations. All fifteen networks reach held-out accuracy $1.000$, with final losses between $1.9\times10^{-5}$ and $3.6\times10^{-4}$; at the level of task performance, they are indistinguishable.

\textit{Phases of inference.}---We classify each network by the fate of the fixed point of the disfavored decision as $h$ increases, tracking the leading nontrivial multiplier after removing the exact symmetry mode. We locate a real unit-multiplier crossing by solving
\[
F_\theta(z;h)=z,\qquad
(J-\mathbb{I})v=0,\qquad
v^{\top}v=1,
\]
and identify a network as Neimark--Sacker only when no fold is found by the extended system, the fixed-point branch persists, and a complex-conjugate pair crosses the unit circle~\cite{SM}. Three principal phases emerge. In the \emph{fold} phase, a real multiplier reaches unity as the fixed point collides with a saddle and both disappear. In the \emph{Neimark--Sacker} phase, a complex-conjugate pair crosses the unit circle, destabilizing the fixed point while leaving bounded nonstationary dynamics. In the \emph{rigid} phase, the bistable branch persists throughout the explored field range. Monostable and non-generic cases occur as additional outcomes~\cite{SM}.

\begin{figure*}[t]
\includegraphics[width=0.9\textwidth]{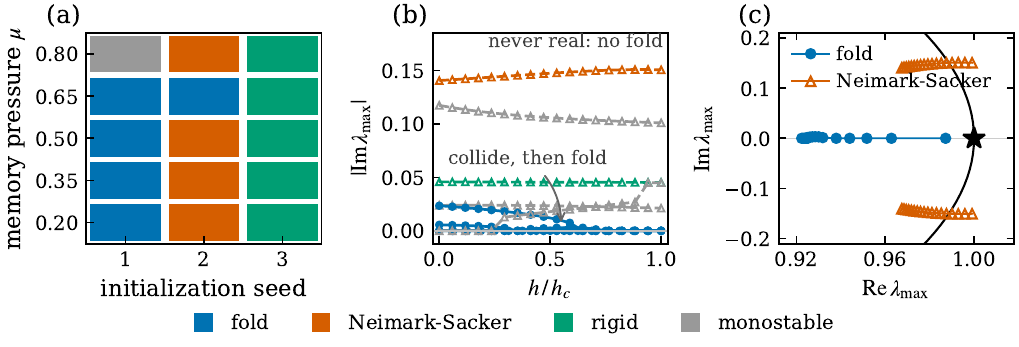}
\caption{\label{fig1}(a)~Phase across memory pressure and initialization, fifteen networks. (b)~$|\mathrm{Im}\,\lambda_{\max}|$ along the branch, with the symmetry mode removed; fold networks are solid. (c)~$\lambda_{\max}$ in the complex plane for one fold and one Neimark--Sacker network.}
\end{figure*}

Figure~\ref{fig1}(a) summarizes the phase assignments across memory pressure and initialization. The phase is strongly initialization-dependent but only weakly dependent on the objective. The phase differs across seeds at every value of $\mu$, whereas varying $\mu$ at fixed initialization changes the classification only twice across the sweep. At fixed $\mu=0.50$, ten independent initializations yield five folds, one Neimark--Sacker, one rigid, and three monostable networks, all with accuracy $1.000$~\cite{SM}. Checkpoints further show that the phase becomes identifiable early in training: eight of ten networks are already in their final phase by step $250$; one additional network subsequently changes locally from a non-generic crossing to Neimark--Sacker, while none switches directly between fold and Neimark--Sacker, despite a further $4$--$24$-fold reduction in loss after the phase is established~\cite{SM}. Thus, task performance does not resolve the dynamical distinction: training finds multiple solutions with essentially identical objective values but qualitatively different inference dynamics.

\textit{What symmetry constrains.}---LayerNorm is invariant under
$z\to z+c\bm1$, so a perturbation along $\bm1$ survives normalization only through the residual decay, giving exactly
\begin{equation}
J\bm1=(1-\gamma)^{2}\bm1
\end{equation}
at every state and field~\cite{SM}. This symmetry mode carries no information about the solution and is therefore removed before the remaining spectrum is analyzed; otherwise $(1-\gamma)^2$ can appear among the leading multipliers away from the fold. The Jacobians of the nonlinearities are symmetric, so antisymmetry, and hence rotation, enters through compositions of the learned weight matrices. Looped recurrences are known to exhibit substantial non-normality and rotation even at random initialization~\cite{buehlmaier2026}; here the decomposition shows that their magnitude and direction vary across initializations of the same architecture.

These constraints narrow the bifurcations observed in our sweep. Because the map is real, multipliers are real or occur in conjugate pairs. No architectural symmetry protects a pitchfork, and we find no transcritical crossings; period doubling is also absent, with the closest approach satisfying $|\lambda+1|=1.7214$. The observed codimension-one bifurcations are therefore folds and Neimark--Sacker crossings, apart from occasional non-generic crossings~\cite{SM}. This is a classification of the models explored, not an exclusion theorem.

The decisive quantity is not the spectrum at the resting attractor but its evolution with the control field. Resting-state rotation does not separate the phases: the antisymmetric fraction of the Jacobian overlaps between the fold and Neimark--Sacker networks, spanning $[0.067,0.140]$ and $[0.098,0.190]$, respectively~\cite{SM}. Likewise, a small imaginary part at rest does not preclude a fold, because a complex pair can collide on the real axis as $h$ increases. We therefore classify the phase from the continued Jacobian spectrum rather than from any scalar descriptor at rest. What matters is not how much rotation the map carries, but whether it carries rotation in the critical direction.

\textit{Inside the fold phase, an effective theory with no fitted parameters.}---At a generic fold, one multiplier reaches unity while the remaining multipliers stay inside the unit circle. The center manifold theorem for maps~\cite{guckenheimer1983,kuznetsov2004} therefore reduces the dynamics to
\[
u_{t+1}=u_t+g\varepsilon+bu_t^2,
\qquad
\varepsilon=h-h_{sn},
\]
where $u=l^{\top}(z-z_{sn})$ and
\begin{equation}
g=l^{\top}\partial_hF_\theta,
\qquad
b=\tfrac12\,l^{\top}\partial^{2}F_\theta[r,r],
\end{equation}
with $l^{\top}r=1$. Both coefficients are determined directly from the trained map; choosing the orientation such that $b>0$ gives $g>0$ for every fold network~\cite{SM}.

\begin{figure*}[t]
\includegraphics[width=0.9\textwidth]{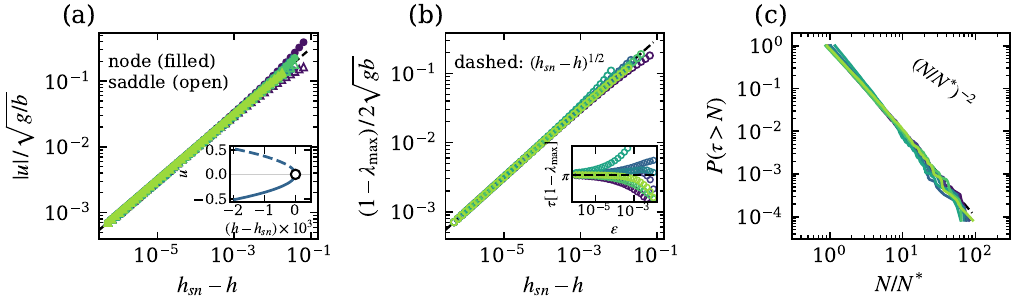}
\caption{\label{fig2}Seven fold-phase networks, each rescaled by its predicted amplitude; dashed lines carry no free parameter. (a)~Node (filled) and saddle (open) branches; inset, sign retained. (b)~Critical slowing down; inset, $\Pi(\varepsilon)$ of Eq.~(\ref{eq:pi}). (c)~Complementary cumulative distributions of $\tau$.}
\end{figure*}

For $\varepsilon>0$, the continuum limit gives $\dot u=g\varepsilon+bu^2$, the normal form of Type-I intermittency~\cite{pomeau1980}. Passage an equal distance $\varepsilon$ above the fold and critical slowing on the stable branch an equal distance below it give, respectively,
\begin{equation}
\tau(\varepsilon)=\frac{\pi/2}{\sqrt{gb}}\,\varepsilon^{-1/2},
\qquad
1-\lambda_{\max}(-\varepsilon)=2\sqrt{gb}\,\varepsilon^{1/2}.
\label{eq:pair}
\end{equation}
The factor $\pi/2$, rather than $\pi$, reflects the one-sided passage from the fold state; the finite-threshold correction is below $0.15\%$ at $\varepsilon=10^{-6}$~\cite{SM}. Thus the exponents and amplitudes are fixed by the same two local derivatives. Evaluating the two relations at equal distances on opposite sides of the fold eliminates $g$ and $b$:
\begin{equation}
\Pi(\varepsilon)\equiv
\tau(\varepsilon)
\big[1-\lambda_{\max}(-\varepsilon)\big]
=\pi+O(\varepsilon^{1/2}).
\label{eq:pi}
\end{equation}
Unlike the exponents, which follow directly from the normal form, this amplitude relation provides a parameter-free test of the reduction itself. A map could reproduce the exponents while failing this amplitude relation. Across seven fold-phase networks spanning $d=16$--$20$, $K=24$ and $48$, both protocols, and $h_{sn}=0.021$--$0.215$, all predictions agree with the measurements to within a few percent~\cite{SM}; Fig.~\ref{fig2} and Table~\ref{tab1} summarize the tests.

\begin{table}[b]
\caption{\label{tab1}Parameter-free tests over seven fold-phase networks spanning $d=16$--$20$, $K=24$ and $48$, both protocols, and $h_{sn}=0.021$--$0.215$. Uncertainties are standard deviations across networks~\cite{SM}.}
\begin{ruledtabular}
\begin{tabular}{lcc}
quantity & predicted & measured\\
\hline
branch exponent & $1/2$ & $0.4999\pm0.0014$\\
slowing exponent & $1/2$ & $0.5004\pm0.0019$\\
slowing amplitude $/\,2\sqrt{gb}$ & $1$ & $1.0016\pm0.0082$\\
passage exponent & $-1/2$ & $-0.4997\pm0.0092$\\
passage prefactor $/\,\tfrac{\pi/2}{\sqrt{gb}}$ & $1$ & $1.0063\pm0.0255$\\
$\Pi$ at $\varepsilon=10^{-6}$ & $\pi=3.1416$ & $3.146\pm0.017$\\
tail exponent, least squares & $-2$ & $-2.130\pm0.196$\\
tail exponent, Hill & $2$ & $2.108\pm0.183$\\
\end{tabular}
\end{ruledtabular}
\end{table}

\textit{The cost of computation follows the phase.}---A budget is spent across a population of problems. For a difficulty density $p$ that is smooth and nonzero at the fold, inverting the first relation in Eq.~(\ref{eq:pair}) gives
\[
p(\tau)\sim 2A^2p(0)\tau^{-3},
\qquad
A=\frac{\pi/2}{\sqrt{gb}},
\]
and hence
\begin{equation}
P(\tau>N)\sim N^{-2}.
\end{equation}
The universality is that of the fold normal form combined with the regularity of the workload density, not of the network itself. For the uniform ensembles used here, $\varepsilon\sim\mathcal{U}(0,\varepsilon_{\max})$ and $p(0)=1/\varepsilon_{\max}$, fixing the amplitude as
\[
P(\tau>N)=\left(\frac{N^*}{N}\right)^2,
\qquad
N^*=A\varepsilon_{\max}^{-1/2}.
\]
Across $8.4\times10^4$ problems in seven networks, with no censoring, the distributions collapse onto $(N/N^*)^{-2}$ with unit amplitude. At $N=10N^*$, the predicted unsolved fraction $10^{-2}$ is measured as $0.0070$--$0.0110$ [Fig.~\ref{fig2}(c)].

For this decision task, $P(\tau>N)$ is the residual fraction of problems unsolved at budget $N$ and therefore the budget-induced error if an unresolved trajectory is counted as an error. The tail also implies
\[
\langle\tau\,|\,\tau>N\rangle\sim2N,
\]
so the expected remaining compute remains of the same order as that already spent. Moreover, $\langle\tau\rangle$ is finite while $\langle\tau^2\rangle$ diverges logarithmically with the cutoff, placing the workload at the boundary between finite and divergent compute fluctuations: rare hard instances dominate the variance.

\begin{figure}[b]
\includegraphics[width=0.9\columnwidth]{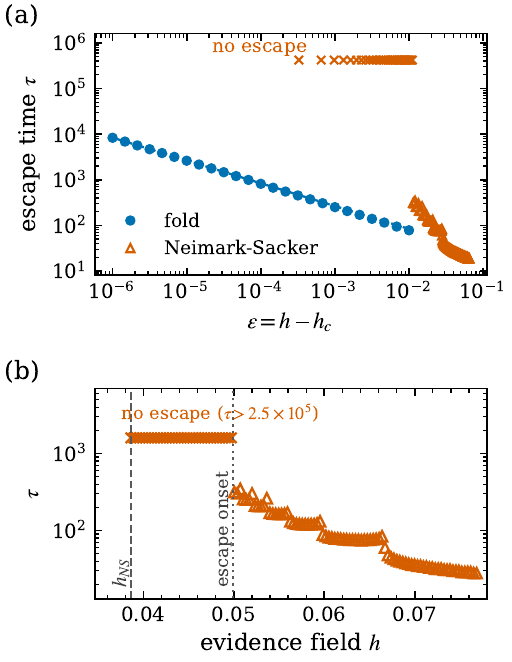}
\caption{\label{fig3}(a)~Escape time against distance to the critical field for a fold network (blue; dashed, Eq.~(\ref{eq:pair})) and a Neimark--Sacker network (orange); crosses mark fields with no escape within $2.5\times10^{5}$ iterations. (b)~The Neimark--Sacker scan on a linear field axis. The non-escaping fields form a contiguous low-field interval and escape begins at a single onset near $h=0.0499$, well above the crossing $h_{NS}=0.0386$: loss of fixed-point stability and escape are distinct events.}
\end{figure}

\textit{In the Neimark--Sacker phase, the fold law disappears.}---Changing phase does not merely change the prefactor of the fold law; it removes the underlying fixed-point-annihilation mechanism. In the Neimark--Sacker network, the disfavored fixed point is not destroyed: a complex-conjugate pair crosses the unit circle at $h_{NS}=0.0386$ with multiplier $0.9886+0.1504i$, an angular increment of $0.151$ radians per iteration, or about $42$ iterations per turn. Past the crossing the fixed point remains present but unstable, and trajectories launched from the branch state remain bounded and nonstationary over the integration interval~\cite{SM}. Scanning $200$ fields above $h_{NS}$, the $35$ non-escaping fields form one contiguous low-field interval ending at $h=0.0497$, followed by a single transition into the escaping regime [Fig.~\ref{fig3}]. Escape therefore begins at a field $29\%$ above $h_{NS}$, well separated from the crossing, and we assign no fold-like power-law exponent to this phase. The fixed point persists, and escape is controlled by the subsequent loss of the bounded nonstationary state rather than by passage through a saddle-node bottleneck. Thus the fold-derived compute-time law has no direct analogue in this network. We analyze escape only for this phase; the rigid and monostable networks have no fold at which the same slowing law can arise. Predictability of cost is itself a property of the dynamical phase.

\textit{Discussion.}---That a trained network organizes inference around a bifurcation is not surprising; attractor models of decision making have been built this way for decades~\cite{wong2006,wang2008,mante2013,sussillo2013}, and networks trained on a single task can realize it through distinct mechanisms whose selection depends on initialization~\cite{ghazizadeh2021,jarne2023,jarnelaje2023,turner2021}. What is new here is that these alternatives are distinct bifurcation phases, and that the selected phase determines the computational time scale. This sharpens rather than contradicts the observation that trained recurrent networks can share dynamical topologies~\cite{maheswaranathan2019}: such sharing can hold within a phase even when the phase itself is not fixed by the task. It also extends the implicit bias identified for looped \emph{linear} transformers with normalization~\cite{wu2026}: optimization can favor a dynamical class, but that class need not be determined by the task.

From a physics perspective, optimization acts as a quench that selects among dynamically inequivalent solutions that are indistinguishable at the level of task performance, with symmetry constraining the available structure rather than selecting the phase. Once selected, the phase admits a remarkably sharp description. In the fold phase, a nonlinear map produced by gradient descent, whose objective contains no bifurcation or compute-time constraint, is quantitatively captured by the saddle-node normal form: both critical exponents and amplitudes are determined by local derivatives of the trained map, including the parameter-free relation of Eq.~(\ref{eq:pi}). This is normal-form universality, arising from the structural stability of the fold under smooth coordinate changes, rather than fluctuation-driven universality of a thermodynamic transition. There is no diverging correlation length or thermodynamic limit; the exponent $1/2$ follows directly from the local Taylor expansion.

The fold normal form further converts single-problem critical slowing down into an ensemble theory of compute. Combining $\tau\sim\varepsilon^{-1/2}$ with a workload density regular at the bottleneck gives $P(\tau>N)\sim N^{-2}$, placing the workload at the boundary of finite compute fluctuations. The ingredients are the local normal form and the regularity of the difficulty distribution, not the details of the neural network. The same composition should therefore apply to other iterative computations whose instances approach a bottleneck with varying distance to criticality. By contrast, the Neimark--Sacker phase lacks a fixed-point annihilation and, in our data, no analogous single-parameter critical compute-time scaling emerges. Predictability of test-time compute is therefore itself a property of the dynamical phase.

This perspective also clarifies the relation to stability analyses. Previous work asks whether the recurrence settles and shows that a spectral bound on the linear part does not by itself answer that question~\cite{prairie2026,buehlmaier2026}; we ask what the full nonlinear map does as it approaches a bifurcation, and how the resulting relaxation time scales. The non-settling trajectories reported at random initialization~\cite{buehlmaier2026} are consistent with the nonstationary dynamics of our Neimark--Sacker phase, but our results show that such dynamical alternatives can persist after training and depend strongly on initialization. This also reconciles the saturating exponential observed in test-time curves~\cite{prairie2026}, associated with the spectral radius of the full Jacobian~\cite{buehlmaier2026}, with the power-law workload tail found here: exponential relaxation describes the dynamics of a fixed problem, whereas the power law arises from the distribution of relaxation times across problems. The two describe different levels of the computation and are therefore not in conflict.

Several predictions follow that can be tested on existing looped models using per-instance iteration counts~\cite{SM}. If a workload samples a fold bottleneck for which the leading real multiplier approaches unity and the difficulty density is regular there, the iteration counts should develop the $N^{-2}$ tail even if mean performance retains an approximately exponential dependence on compute. More generally, if the difficulty density behaves as $p(\varepsilon)\sim\varepsilon^{-a}$ near the bottleneck, the same composition gives $P(\tau>N)\sim N^{-2(1-a)}$, or a positive tail index $\theta=2(1-a)$ when $P(\tau>N)\sim N^{-\theta}$; thus the measured tail probes the distribution of distances to criticality. Architectures with $\rho(\overline{\bm A})=1$ by construction~\cite{prairie2026} provide a natural setting for this test, although marginal linear residual dynamics does not by itself imply a fold. Finally, within our bistable ensemble, every network whose leading nontrivial multiplier was exactly real at the resting attractor belonged to the fold phase, although not every fold was real at rest. Since the $N^{-2}$ law was observed only in the fold phase, the resting spectrum is a candidate diagnostic whose scope must be tested beyond the present ensemble.

\textit{Conclusion.}---Training determines what a looped transformer computes, while spectral conditions on its linearized recurrence constrain the stability of its residual dynamics. Neither, however, determines how inference unfolds. That is set by the dynamical phase selected by training: networks with the same architecture, objective, and task performance can relax through qualitatively different bifurcations, with the choice depending strongly on initialization and becoming identifiable when the memory--attractor structure forms. Symmetry constrains the available phases but does not select among them. In the fold phase, the resulting dynamics obey a parameter-free critical scaling that extends from single-problem relaxation to a universal workload tail, whereas the Neimark--Sacker phase has no analogous compute-time law. Test-time compute is therefore not determined by the architecture alone; it is a dynamical property of the solution that training finds.

\textit{Data availability.}---The data and code that support the findings of this study are available from the corresponding author upon reasonable request.

\end{document}